\documentclass[sigconf]{acmart}

\AtBeginDocument{%
  \providecommand\BibTeX{{%
    \normalfont B\kern-0.5em{\scshape i\kern-0.25em b}\kern-0.8em\TeX}}}

\usepackage{soul}
\usepackage{color}
\usepackage{wallpaper}
\usepackage{graphicx}
\usepackage{cleveref}
\usepackage{numprint}
\definecolor{orangeX}{rgb}{1,.5,0}
\definecolor{blueX}{rgb}{.2, .59, .88}
\definecolor{purpleX}{rgb}{.294118, 0, .509804}
\definecolor{greenX}{rgb}{.721, .878, .341}
\definecolor{bole}{rgb}{0.47, 0.27, 0.23}
\definecolor{mypink3}{cmyk}{0, 0.7808, 0.4429, 0.1412}
\definecolor{mygray}{gray}{0.6}
	
\newcommand{\musec}{MuSe-CaR\,}
\newcommand{\musewild}{MuSe-Wild\,}
\newcommand{\musetopic}{MuSe-Topic\,}
\newcommand{\musetrust}{MuSe-Trust\,}
\newcommand{\ds}{\textsc{DeepSpectrum\,}}
\newcommand{\opensmile}{\textsc{openSMILE\,}}
\newcommand{\openface}{\textsc{OpenFace\,}}
\newcommand{\egm}{\textsc{eGeMAPS\,}}
\newcommand{\ft}{\textsc{FastText\,}}
\newcommand{\vgg}{\textsc{VGGface\,}}
\newcommand{\xce}{\textsc{Xception\,}}
\newcommand{\lld}{\textsc{low-level descriptors\,}}
\newcommand{\go}{\textsc{GoCaR\,}}
\newcommand{\op}{\textsc{OpenPose\,}}
\newcommand{\fau}{\textsc{Facial Action Units\,}}
\newcommand{\mtcnn}{\textsc{MTCNN\,}}

\newcommand{\eg}{e.\,g.,\,}
\newcommand{\ie}{i.\,e.,\,}

\newcommand{\cf}{{cf.\,}}

\begin{document}

\title{MuSe 2020 -- The First International \\Multimodal Sentiment Analysis \\ in Real-life Media Challenge and Workshop}
\subtitle{Emotional Car Reviews in-the-wild}

\author{Lukas Stappen}
\affiliation{%
  \institution{University of Augsburg}
  \city{Augsburg, Germany}}

\author{Alice Baird}
\affiliation{%
  \institution{University of Augsburg}
  \city{Augsburg, Germany}}

\author{Georgios Rizos}
\affiliation{%
  \institution{Imperial College London}
  \city{London, UK}}
  
\author{Panagiotis Tzirakis}
\affiliation{%
  \institution{Imperial College London} 
  \city{London, UK}}

\author{Xinchen Du}
\affiliation{%
  \institution{Technical University of Munich}
  \city{Munich, Germany}}
  
\author{Felix Hafner}
\affiliation{%
  \institution{University of Augsburg}
  \city{Augsburg, Germany}}
  
  \author{Lea Schumann}
\affiliation{%
  \institution{University of Augsburg}
  \city{Augsburg, Germany}}
  
\author{Adria Mallol-Ragolta}
\affiliation{%
  \institution{University of Augsburg}
  \city{Augsburg, Germany}}

\author{Bj\"orn W. Schuller}
\affiliation{%
  \institution{Imperial College London}
  \city{London, United Kingdom}}

\author{Iulia Lefter}
\affiliation{%
  \institution{Delft University of Technology}
  \city{Delft, Netherlands}}
  
\author{Erik Cambria}
\affiliation{%
  \institution{Nanyang Technological University}
  \city{Singapore}}  

\author{Ioannis Kompatsiaris}
\affiliation{%
 \institution{CERTH - ITI}
 \city{Thermi-Thessaloniki, Greece}
}

\renewcommand{\shortauthors}{Stappen, et al.}

\begin{abstract}
\textbf{Mu}ltimodal \textbf{Se}ntiment Analysis in Real-life Media (MuSe) 2020 is a Challenge-based Workshop focusing on the tasks of sentiment recognition, as well as emotion-target engagement and trustworthiness detection by means of more comprehensively integrating the audio-visual and language modalities. The purpose of MuSe 2020 is to bring together communities from different disciplines; mainly, the audio-visual emotion recognition community (signal-based), and the sentiment analysis community (symbol-based). We present three distinct sub-challenges: \musewild, which focuses on continuous emotion (arousal and valence) prediction; \musetopic, in which participants recognise 10 domain-specific topics as the target of 3-class (low, medium, high) emotions; and \musetrust, in which the novel aspect of trustworthiness is to be predicted. In this paper, we provide detailed information on \musec, the first of its kind in-the-wild database, which is utilised for the challenge, as well as the state-of-the-art features and modelling approaches applied. For each sub-challenge, a competitive baseline for participants is set; namely, on test we report for \musewild a combined (valence and arousal) CCC of $.2568$, for \musetopic a score (computed as 0.34$\cdot$ UAR + 0.66$\cdot$F1) of $76.78$\,\% on the 10-class topic and $40.64$\,\% on the 3-class emotion prediction, and for \musetrust a CCC of $.4359$. 
\end{abstract}




\maketitle

\section{Introduction}
\textbf{Mu}ltimodal \textbf{Se}ntiment Analysis in Real-life Media (MuSe) 2020 is a novel Challenge-based Workshop in which \emph{sentiment recognition}, as well as \emph{emotion-target engagement} and \emph{trustworthiness detection} are the main focus. MuSe aims to provide a testing bed for more extensively exploring the fusion of the audio-visual and language modalities. The core purpose of MuSe is to bring together communities from differing computational disciplines; mainly, the sentiment analysis community (symbol-based), and the audio-visual emotion recognition community (signal-based). 

The first group -- rooted in the field of Sentiment (and Opinion) Mining and specialising in Natural Language Processing (NLP) methods for symbolic information analysis -- leverages the text modality, and focuses on the prediction only of discrete sentiment label categories~\cite{zadeh2018proceedings}. 
In numerous competitions from recent years researchers from the second group -- mostly rooted in the field of Affective (and Behavioural) Computing and specialised in intelligent signal processing --  focused on one, or both of the audio and vision modalities, in order to predict the continuous-valued valence and arousal dimensions of emotion (circumplex model of affect), while often disregarding the potential contribution of textual information~\cite{valstar2013avec,ringeval2017avec, kollias2020analysing,schuller2018interspeech}. However, approaches by both communities now show signs of convergence, highly influenced by related, explicitly multimodal learning techniques~\cite{arevalo2020gated, gomez2020exploring, qiu2020multimodal}. Of note, the 2020 INTERSPEECH Computational Paralinguistics (ComParE) Challenge included for the first time baselines utilising both audio signal and text transcripts~\cite{schuller2020interspeech}. 

With this in mind, MuSe 2020 aims to attract both communities equally and encourages a fusion of modalities to demonstrate the advantages within the field of emotion specifically. Ideally,  participation should strive towards the development of unified approaches applicable to each task. Tasks have arisen from different academic traditions: on the one hand, complex, dimensional emotion annotations relating to the expression of behaviour, and on the other hand, linking sentiment and emotions to topics (context), entities or aspects, as is common in sentiment analysis~\cite{soleymani2017survey}.

A second contribution of MuSe 2020 is the facilitation of a broad comparison of the merits for the three core modalities (language, audio, and visual cues), as well as various approaches of multimodal fusion under well-defined and strictly comparable conditions. In this way, establishing the extent to which the fusion of approaches is possible and beneficial, as well as advancing sentiment and emotion recognition systems to be able to deal with fully naturalistic (in-the-wild) behaviour from large volumes of in-the-wild (user-generated) data. User-generated data types refers to data sourced from the target user themselves and are the new generation of data utilised for real world multimedia affect and sentiment analysis~\cite{wang2020discovering} and other research fields~\cite{cuomo2020user}.

For all of the three sub-challenges, one dataset is chosen to make the comparison between each sub-challenge more easily facilitated. For this year's MuSe 2020, we introduce the Multimodal Sentiment Analysis in Car Reviews dataset \emph{\musec} which covers the range of aforementioned topics discussed. \musec is a large, multimodal dataset which has been gathered in-the-wild with the intention of further understanding real world Multimodal Sentiment Analysis, in particular the emotional engagement that takes place during product reviews (\ie automobile reviews) where a sentiment is linked to a topic or entity. 

%\ls{@all: If you have suitable related work, incorporate it above since we are not going to have a related work section (space reasons). Thanks!}
% \section{Related Work}

\section{Challenge Outline and Protocol}
The major novelties discussed herein will be introduced in MuSe 2020 through three core sub-challenges, (i) Multimodal Sentiment in-the-Wild Sub-challenge (\musewild), (ii) Multimodal Emotion-Target Engagement Sub-challenge (\musetopic) (iii) Multimodal Trustworthiness Sub-challenge (\musetrust). In the following, we will describe and highlight the aforementioned novelties of each each sub-challenge, as well as include guidelines for participation.  
%BS: PLEASE CHANGE!!! Cumbersome, unclear... Please use Sentiment Target or Emotion Target or Target & Engagement if need be 
%LS: changed naming, but for abbreviation it is unfor. too late since we already sent out data packages including this abbreviation...

Individuals wishing to participate in the MuSe 2020 challenge must hold an academic affiliation. Further to this, they should download and fill out the End User License Agreement (EULA) and submit via the homepage\footnote{www.muse-challenge.org}. All entries to the challenge should be accompanied with a document which describes in detail methods and results and includes a citation of this paper. To appear on the temporary, public leader board on the MuSe homepage, participants must provide predictions, a Github repository where their source code is uploaded, and a link to an arXiv preliminary technical report. The organisers do not participate in the Challenge themselves, but re-evaluate the findings of the best performing system of each Sub-challenge. There will be a double blind peer-reviewed process by the technical program committee, and only papers which meet the standards set by peer-review will be eligible for the main competition. Papers accepted for the workshop will be allocated 6-8 pages (plus %infinite
references) in the proceedings of ACM MM 2020. 
\vspace{-0.2cm}

\subsection{\musewild Sub-Challenge}
In the \emph{\musewild Sub-Challenge}, participants are predicting the level of affective dimensions (arousal, and valence) in a time-continuous manner from audio-visual recordings. Valence thereby is strongly linked to the emotional component of the umbrella term of sentiment analysis and is often used interchangeably \cite{thelwall2010sentiment, mohammad2016sentiment,preoctiuc2016modelling}. Timestamps to enable modality alignment and fusion on word-, sentence-, and utterance-level as well as several acoustic, visual and textual-based features are pre-computed and provided with the challenge package. The evaluation metric for this sub-challenge is \textit{concordance correlation coefficient (CCC)}, which is often used in similar challenges \cite{valstar2013avec, ringeval2017avec}. CCC is a measure of reproducibility and performance, which condenses information on both precision and accuracy, is robust to changes in scale and location \cite{lawrence1989concordance}, and its theoretical properties to other regression measures, \eg (root) mean squared error, are well understood \cite{pandit2019many}. For the baseline for the \musewild sub-challenge the mean of arousal and valence is taken.

\subsection{\musetopic Sub-challenge}
In the \emph{\musetopic Sub-challenge}, participants are predicting 10-classes of domain-specific (automotive, as given by the chosen database) topics\footnote{Classes for the \musetopic sub-challenge include; General Information, Costs, Performance, Quality \& Aesthetic, Safety, Comfort, Exterior Features, Interior Features, Handling/Driving Experience, User Experience.} as the target of emotions. In addition, three classes (low, medium, and high) of valence and arousal should be predicted \ie for each topic segment, one valence and one arousal value. These classes are the mean value of the temporally aggregated continuous labels of \musewild, which were divided into three equally sized classes (33\,\%) for each label For this sub-challenge, first, the weighted score combining (0.34$\cdot$) Unweighted Average Recall (UAR) and (0.66$\cdot$) F1 (micro) measures independently for each predictions (Valence, Arousal and Topic) are calculated. We include both these factors to keep our evaluation consistent with previous challenges, as the former was partially used to evaluate a classification task in \cite{kollias2020analysing}, and the latter in \cite{schuller2020interspeech}. Second, the mean of the weighted scores for Valence and Arousal (combined) is calculated. Third, to combine the mean with the topic score the mean rank over all participants ((rank of combined emotions result + rank of topic result)/2) is calculated for the final performance assessment. In case two participants should have the same mean rank, the one with the highest topic rank will be the final winner. We believe that this composite measure is most discriminative to meaningfully showcase performance improvements in emotion and topic prediction, as it places importance on precision and recall, in both a dataset-wide and class-specific manner. 
\vspace{-0.2cm}

%The score is equal to: $0.66 \cdot \text{Micro-F1} + 0.33 \cdot \text{UAR}$. 
% We select as a performance measure an interpolation of Micro-F1 and Unweighted Average Recall (UAR or Macro-Recall).
%In addition to the provided features from \musewild, a visual domain-specific entity recognition and localisation model enabling an easy, domain-specific utilisation of visual features are provided. 
\vspace{-0.5em}
\subsection{\musetrust Sub-challenge}
In the \emph{\musetrust Sub-challenge}, participants are predicting a continuous trustworthiness signal from user-generated audio-visual content in a sequential manner and are provided with aligned valence and arousal annotations, which participants are encouraged to explore, in a means of understanding the relationship between emotional labels in depth and at large scale. The evaluation metric for this sub-challenge is concordance correlation coefficient (CCC).

\section{Challenge Dataset}
\label{sec:corpus}

\begin{table}[t!]
\footnotesize
  \caption{
Partitioning of the \musec dataset, applied for each of the three sub-challenges. Reported are the number of unique videos, and the duration for each sub-challenge hh\,:mm\,ss. The unprocessed duration of the \musec dataset is 36\,:52\,:08.  %\ls{for \musetopic, should we report the number of segments here because one segment one label not like the others??}
 }
  \begin{tabular}{l|rrrr}
    \toprule
    Partition & No. & \musewild & \musetopic  & \musetrust \\
    \midrule
    Train   & 166 & 22\,:16\,:43 &22\,:35\,:55 & 22\,:45\,:52 \\
    Devel.  &  62 & 06\,:48\,:58 &06\,:49\,:46 & 06\,:52\,:22 \\
    Test    &  64 & 06\,:02\,:20 &06\.:14\,:08 & 06\,:12\,:53 \\
    \hline
    $\sum$    & 291 & 35\,:08\,:01 & 35\,:39\,:49 & 35\,:51\,:07\\
  \bottomrule
\end{tabular}
\label{tab:paritioning}
\vspace{-0.2cm}
\end{table}

\begin{figure*}
    \centering
    \includegraphics[width=0.3\linewidth]{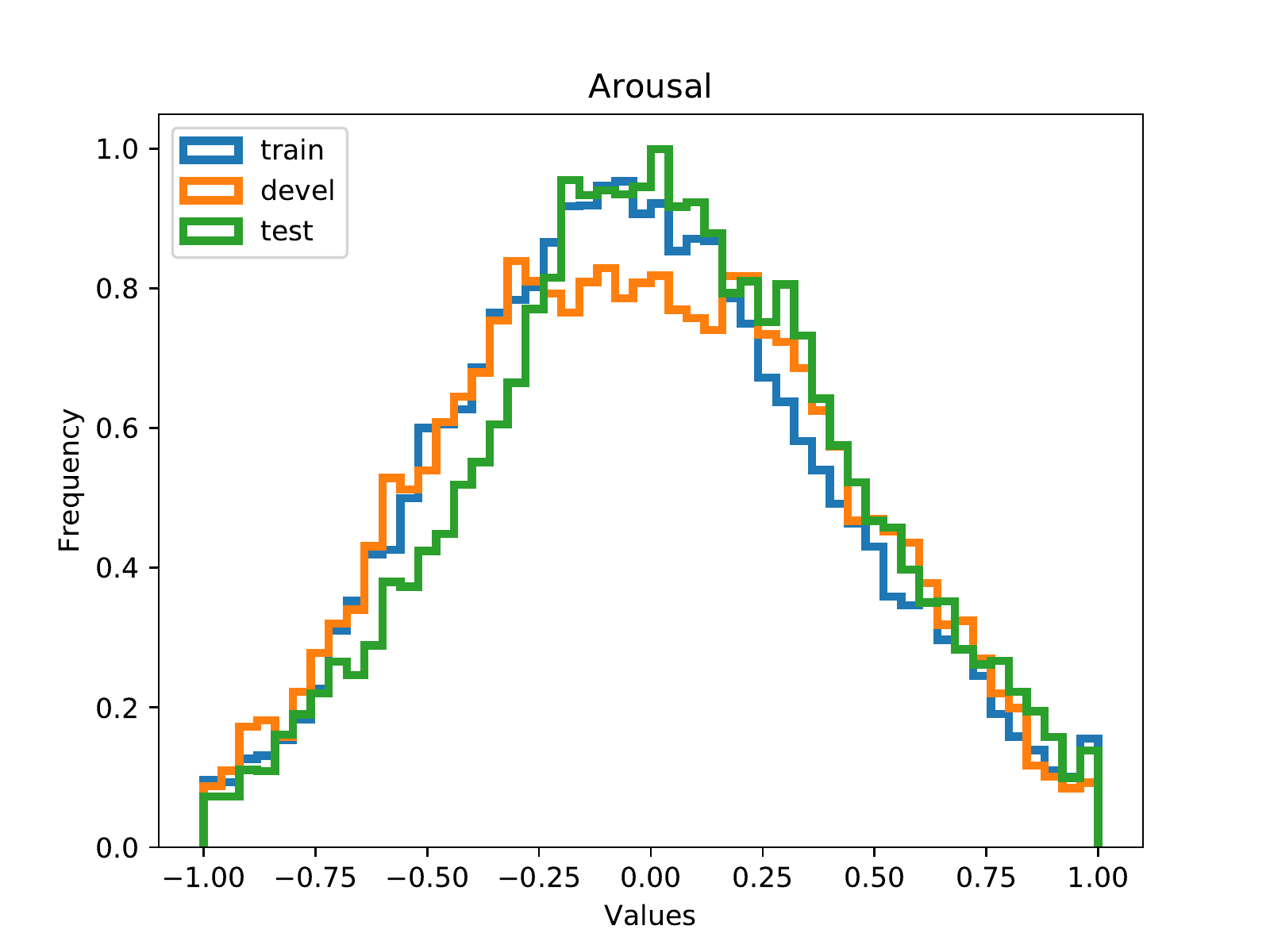}
    \includegraphics[width=0.3\linewidth]{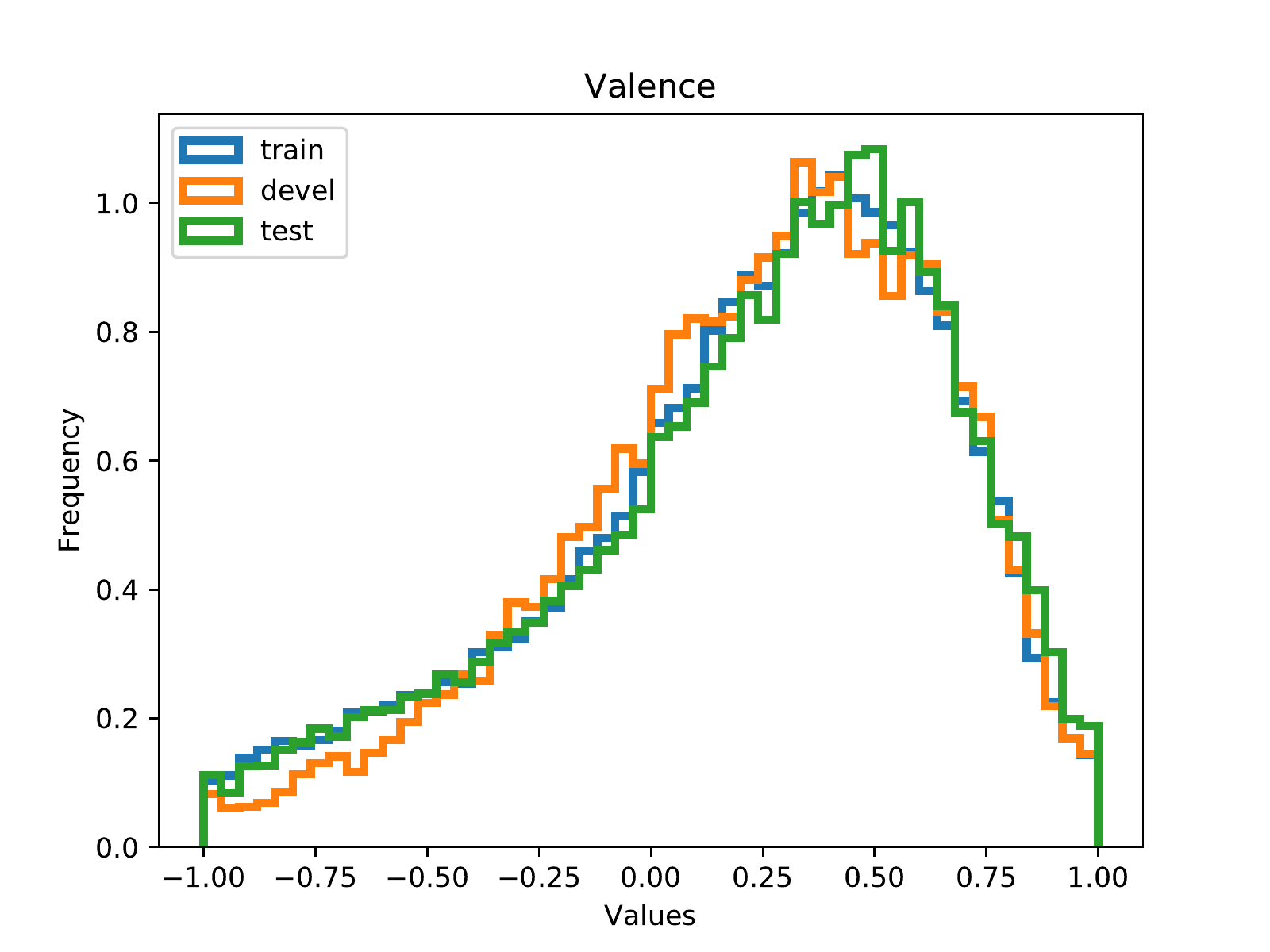}
    \includegraphics[width=0.3\linewidth]{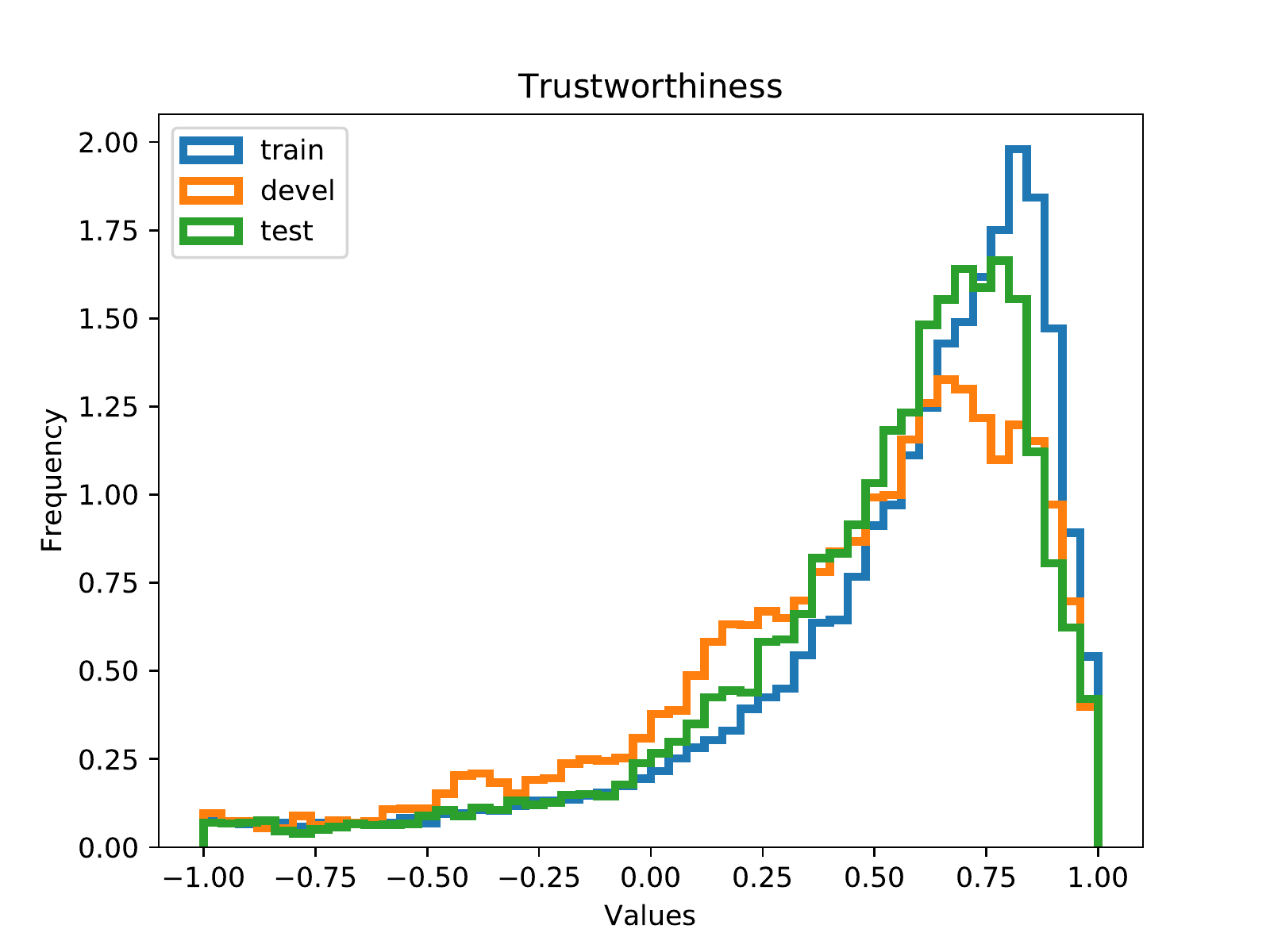} \\
    \vspace{-0.2cm}
    \caption{Frequency distribution in the partitions Train, (Devel)opment, and Test for the continuous prediction sub-challenges \musewild, Arousal and Valence, and \musetrust, Trustworthiness. }
    \label{fig:freq}
    \vspace{-0.4cm}
\end{figure*}
% \ls{train: 4207, devel: 1335, test: 1260}

For all of the three Sub-Challenges of MuSe 2020, the \musec data set is utilised. \musec is a large, extensively annotated multimodal ((spoken) language, audio, video) dataset which has been gathered in-the-wild with the intention of developing appropriate methods and further understanding Multimodal Sentiment Analysis in-the-wild. \musec has been designed with an abundance of computational tasks in mind, including emotion and entity recognition, and dominantly with the intention of improving machine understanding of how sentiment (\ie emotion) is linked to an entity and aspects of such reviews. 

The estimated age range of the professional, semi-professional (`influencers'), and casual reviewers is from the mid-20s until the late-50s. Most are native English speakers from the UK or the US, while a small minority are non-native, yet fluent English speakers. \musec includes high voice and video quality, as everyday recording devices have improved in recent years. This enables robust learning of a high degree of novel, in-the-wild characteristics.

For the MuSe 2020 Challenge, we selected a high-quality sub-set of the \musec dataset consisting of 36\,h\,:\,52\,m\,:\,08\,s of video data from 291 videos and 70 host speakers (plus an additional of roughly 20 narrators) sourced from YouTube. 

When creating the data set, it was of particular importance to find a balance between the stable and uncontrollable, `in-the-wild' properties such as different recording devices, camera perspectives, ambient noises (car noises, music), or changing backgrounds to allow for meaningful learning with current deep learning methods. Such `in-the-wild' characteristics of \musec include; i) \emph{video}: shot size, face-angle,  camera motion, reviewer visibility, reviewer face occlusion (glasses), and highly varying backgrounds; ii) \emph{audio}: ambient noises, narrator and host diarisation, diverse microphone types, and speaker locations; iii) \emph{text}: colloquialisms, and domain-specific terms. 

The topic of videos within \musec is limited to vehicle reviews, with the number of vehicle manufacturers being restricted to premium brands (BMW, Audi, Mercedes-Benz) that equip their vehicles with the latest technology, thus, ensuring that discussed entities and aspects (\eg semi-autonomous vehicle functions) occur across a board range of videos (and different manufacturers). Most of the reviewers are semi- or professional reviewers (\eg YouTube channel `influencers'). All YouTube channels used within \musec have given full consent for their data to be used with the context of academic research\footnote{Following the YouTube guidelines, uploading a video to YouTube automatically issues that video under the YouTube own license. To the best of our understanding, under this licence,  the use of the data in the EU is only possible by YouTube directly or with the consent of the creator. In similar works, the database producers refer to the fair use principle for academic use. Furthermore, the YouTube's standard terms \& conditions as well of those from the YouTube API have to be considered. A fraction of videos are also available under the Creative Commons Licence (CC-BY, full use, if the creator credits are mentioned which are provided in the data packages.)}.

To avoid extremely objective reviews, during the selection process videos were rated on a scale between 0 (emotionless) and 5 (very emotional). We filtered out all videos with a score less than 3 before annotation began.
Within \musec, there are 15 annotation tiers (3 continuous dimensional, 3 partially continuous binary label, 5 categorical, and 4 automatically annotated tiers). For MuSe 2020, we utilise 3 continuous ratings, and the topic categorical ratings. Each recording has been annotated in three continuous dimensions; emotional  \emph{valence} (hence reflecting sentiment) and \emph{arousal} according to Russell’s theory~\cite{russell1980circumplex}, and additionally the novel aspect of \emph{trustworthiness}, each by at least 5 independent annotators. In the case of the Trustworthiness dimension, there has been minimal research into the link between this and other emotions \cite{aguado2011learning}, and to the best of the authors'  knowledge, it has not been utilised nor predicted using machine learning. 

A \emph{gold-standard} was computed on the individual annotators using an Evaluator Weighted Estimator (EWE) approach, in which inter-rater agreement is considered. EWE is described, \eg further in \cite{schuller2013intelligent} and has been applied to similar continuous emotion-based tasks~\cite{ringeval2017avec}, and corpora~\cite{ringeval2013introducing}. %\aebc{do we need to include this here? or/also ccc?}When we apply this approach the mean CC for inter-rater agreement across all data is - arousal: 0.313 ($\pm$ 0.084), valence: 0.415 ($\pm$ 0.093), trustworthiness: 0.413 ($\pm$0.102). 
In addition to the dimensional annotations we included the categorical labelling of emotional engagement with topics, such as comfort, safety, interior, and performance.

For the MuSe 2020 Challenge, data has been partitioned in a Train, Development, and Test convention, where aspects including emotional ratings, speaker independence, and duration have been considered (\cf \Cref{tab:paritioning} for an overview). The total duration of data for each sub-challenge varies, as further pre-processing to include the most informative data only was applied. For \musewild and \musetrust, all parts with an active voice or a visible face are included. We excluded non-product related video segments (\eg advertisements) for \musewild and \musetopic to minimise the distortion these could cause on the task objectives. More specifically, for \musetopic, we only included sections which have an active voice based on the sentence transcriptions. To not fragment it to purely sentence segments, we fused adjacent segments if the segments cover the same topic and are less than two seconds apart. Regarding \musetrust, non-product related information -- for instance, advertisement --, might have a notable impact on the trustworthiness perception of the video. Therefore, segments containing advertisements for products and YouTube channels are included.

\section{Baseline Features}
For each Sub-challenge, we provide a selection of features to participants which have been extracted from language, audio (including speech-to-text), and video signals. Extracting rich features from a huge amount of video data takes days, sometimes weeks, to complete, which would cost participants valuable time. For this reason, we provide $14$ model-ready audio, visual, and linguistic feature sets\footnote{Note: The participants are also free to use external data, any commercial or academic feature extractors, pre-trained networks, and libraries. However, this should be reproducible and clearly discussed in the accompanying paper.}, an amount which far exceeds the number of feature sets provided by other comparable audio-visual challenges \cite{valstar2013avec, ringeval2017avec, kollias2019deep, zadeh2018proceedings}. In the proceeding section,  the feature sets based on each modality (acoustic, vision, and language) are described. For all feature sets a hop size of 0.25\,s was applied (unless otherwise stated) to be inline with the annotation sampling rate. 

\vspace{-0.2cm}
\subsection{Acoustic}
For extracting acoustic features, we utilise well-known feature extraction tools, namely \opensmile and \ds, which have both shown success in a variety of audio processing tasks, including prominent work in speech emotion recognition (SER)~\cite{schuller2013interspeech,cummins2017image}. Audio is extracted directly from the YouTube videos, normalised to -3dB and converted from stereo to mono, 16\,kHz, 16\,bit. For all acoustic features, we apply a window size of 5 seconds. 

\vspace{-0.1cm}
\subsubsection{\opensmile}
The freely available \opensmile toolkit~\cite{eyben2010opensmile} is utilised to extract the well-known extended Geneva Minimalistic Acoustic Parameter Set (\egm)~\cite{eyben2015geneva}. \egm is a hand-crafted speech-based feature set, containing 88 features designed specially for Speech Emotion Recognition (SER) tasks \cite{stappen2019speech}. In addition, 130 dimensional low-level descriptors are provided which have been computed with \opensmile, and include the features 1st and 2nd-order derivatives (deltas and double-deltas). LLD extraction remained at the the default \opensmile configuration and therefore a window size of 10\,ms is applied for this feature set only. 

\vspace{-0.1cm}
\subsubsection{\ds}
We also include \ds features as a state-of-the-art deep learning based approach~\cite{amiriparian2017snore}. \ds features extract spectral images from speech instances and are then fed into pre-trained image recognition Convolutional Neural Networks (CNNs), and the resulting activations are extracted as feature vectors. For MuSe 2020, we extract features utilising the VGG-19 \emph{extraction network}~\cite{simonyan2014very}, with all other parameters remaining as default. This results in a feature set of 4\,096-dimensions.

\vspace{-0.2cm}
\subsection{Vision}
Most visual feature extractors are either designed to localise and extract specific image characteristics and sections (\eg face), or to learn general discriminatory features for classifying (multi-class, multi-label) a large number of images into many classes (ImageNet). We provide participants with raw data (extracted faces), features focusing on human behavior (face, poses) as well as feature sets which capture the environment as a whole (\xce) or the interaction object car (\go). %These allow a comprehensive exploration of the visual perception on multiple levels.
 
\vspace{-0.15cm}
\subsubsection{\mtcnn}
To extract and localise the faces in the videos, 
\mtcnn \cite{zhang2016mtcnn} was used. Internally, it has a cascaded structure of three stages to predict face and landmark position operating in real time. The model is trained on the data sets WIDER FACE \cite{DBLP:journals/corr/YangLLT15b} and CelebA \cite{liu2015faceattributes}. It also provides a confidence measure which allows the false positive to false negative rate to be tuned. Because the frameworks that extract more detailed face features do not provide features for false positives, we chose not to tune the confidence threshold. For the quantitative performance analysis, we labelled a small selection of videos from each channel by hand, and calculated the intersection over union. Depending on the size of the overlap and intersection, we classified the detected bounding boxes into true and false positives. The detector achieved an accuracy of $90$\,\%, and an F1 score of $86$\,\% on the selection of \musec. In addition, we visually inspected the bounding boxes to control the qualitative performance. Both performances underline the very good quality of \mtcnn face extractions. These extractions were used as inputs for \vgg and \openface.

\vspace{-0.15cm}
\subsubsection{\vgg}
\vgg \cite{Omkar2015recognition} is used to extract facial features from the cropped faces that were detected by \mtcnn. Originally intended for face recognition tasks, it outputs a feature vector of size 512 when the top layer is removed. Its main advantage is the comparable performance to other face recognition models while using less data for training. The data set used to train the deep CNN, called VGG16 \cite{simonyan2014very}, is \vgg, collected by the visual geometry group of Oxford. It contains more than 2 500 identities and 2.6 million faces. While consisting of fewer identities and pose/age variations in comparison to its successor \cite{Cao2017vggface}, the number of images is similar in scale. Compared to OpenFace, these features can be used to extract more raw facial features, \eg to learn predictive facial movements from scratch.

\vspace{-0.15cm}
\subsubsection{OpenFace}
Facial features were also extracted from the cropped faces detected with \mtcnn using \openface~\cite{Baltruvsaitis16-OAO}. This toolkit provides a wide range of facial features. We extracted facial landmarks in both 2D (136 features) and 3D (204 features), 6 head pose features, 288 gaze positions, and the intensity and presence of 17 Facial Action Units (FAUs) each for the left side and centre.

\vspace{-0.1cm}
\subsubsection{\xce}
We use \xce \cite{he2016deep} to provide features that capture the environment. \xce is a very deep, state of the art network using residual blocks which enable easier optimisation for large networks. This architecture won the 1st place on the ILSVRC 2015 classification task and other challenges. It is commonly used as feature extractor for general vision features. To obtain the deep representations, we extract the output of the last fully connected layer from the pre-trained \xce network. As a result, a 2048-dimensional deep feature vector is provided for each frame.

\vspace{-0.1cm}
\subsubsection{\go}
\go \cite{stappen2020go} is an domain-specific visual feature extractor enabling the localisation of 28 car parts, such as, door, steering wheel, headlights, and infotainment with which the reviewer interacts inside and outside the vehicle. It is based on a modified YoloV3 framework \cite{redmon2018yolov3} with a Darknet-53 as backbone and is trained with a multi-label, multi-class real-world data set containing 15\,003 vehicle images of 18 different BMW models with up to 100 different feature variants, each. The coverage of a high number of feature variants is necessary to learn robust features, since cars have one of the highest possible number of product variants, e.\,g., the number of Mercedes E-Class equipment variations exceeds the order of $10^{24}$ \cite{pil2004linking}. The extractor achieves a mean average precision of 67.57\,\% ranging from 94\,\% for very distinctive parts such as grills to 14\,\% for less distinctive ones (\eg roof window) on 1\,000 extracted and manually labelled \musec video frames. The provided \go features are converted into an array of fixed size. For this purpose, we use the 10 objects with the highest confidence, and for each object we store the class (one-hot encoded), the confidence and the localisation coordinates (x, y, width, and height). In total, this results in a feature vector of $10*(27+7)$-dimensions.

\vspace{-0.1cm}
\subsubsection{\op} We extracted 18 2D pose keypoints\footnote{The keypoints are: Nose, Neck, Right/Left Shoulder, Right/Left Elbow, Right/Left Wrist, Right/Left Hip, Right/Left Knee, Right/Left Ankle, Right/Left Eye, and Right/Left Ear.} using the method proposed in \cite{8765346}, which yielded the best results in the COCO 2016 keypoints challenge \cite{lin2014microsoft}. We assume that at maximum, only one person is present in each frame. We use the pre-trained model provided by the authors in \cite{8765346}, trained on the COCO 2016 dataset \cite{lin2014microsoft}. The model consists of two branches of stacked CNNs, where one predicts 2D confidence maps for the keypoints of interest, and the other predicts Part Affinity Fields that contain information on the association of keypoints of the same individual amongst themselves. At each level, the outputs of each branch are concatenated and given as input to the higher level layer pair. In the end, we provide the 2D coordinates, as well as the corresponding confidence value of a keypoint being present, for each of the 18 keypoints.

\subsection{Language}

\ft \cite{bojanowski2017enriching} is a library for efficient learning of word embeddings. It is based on the skipgram model where a vector representation is associated to each character n-gram. The model is trained on the English Common Crawl corpus (600B tokens). In comparison to other traditional word embeddings, such as, word2vec \cite{mikolov2013distributed}, or GloVe \cite{pennington2014glove}, these sub-words chunks make it possible to calculate word representations of words which were not part of the original training corpus (out-of-vocabulary).This appears advantageous since we work with a domain-specific corpus including technical terms and model names. This a valuable function, and enables us to transform 96\,\% of words to word embedding vectors. %In \cite{joulin2017bag} they propose a simple and efficient baseline for text classification. Even though it is faster for training and evaluation, it shows competitive performance.

\vspace{-0.2em}
\subsection{Alignment}
The wide diversity of feature types from three modalities and the correspondingly different sampling rates lead to different lengths of the extracted features along the time axis. All continuous visual feature extractors (e.\,g., \xce, \go) sample 4 frames per second, which corresponds approximately to the 250\,ms labeling and the 250\,ms audio sampling of \ds and \egm (except low-level descriptors which are sampled every 10\,ms). Furthermore, Human-focused features (\eg \vgg, \fau) are only extracted from the frames when a reviewer is visible. Recent work \cite{yao2019MultT} has shown that even when advanced alignment mechanisms are in-cooperated in a multimodal neural network, such as attention heads models, the nets are more effective when the features are first aligned during pre-processing. Therefore, we provide for each sub-challenge non-aligned, label-aligned, and, additionally for the more text-related task \musetopic, \ft-aligned features. If desired, the non-aligned features can be aligned by the participants using the corresponding timestamps (or start and end time of a segment for \musetopic). The label-aligned features have exactly the same length (and timestamps) as the provided label files. We applied zero-padding to the frames, where the feature type is not present or which prevented the extraction of features under unfortunate conditions, e.\,g., \textsc{\openface} when no face appears or when only small faces appear in the original frame.
Only the \ft features are repeated for the duration of a word and non-linguistic parts are also imputed with zero vectors. For the \ft alignment, the features are aggregated in such a way that for a \ft feature vector only one corresponding aggregated feature of any other type exists. This preparation should enable the participants to get started quickly and at the same time allows for own imputation procedures as well as unaligned modelling.

\vspace{-0.1cm}
\section{Baseline Systems}

For each Sub-challenge, a series of state-of-the-art approaches have been applied, and for reproducibility, all resources are made freely available \footnote{https://github.com/lstappen/MuSe2020}. 
In the proceeding section, we describe in detail the approaches. An overview of all baseline results is given in \Cref{tab:base13}, \Cref{tab:base2a}, and \Cref{tab:base2b}. For both Sub-Challenges \emph{\musewild}, and \emph{\musetrust}, the paradigm is continuous prediction of emotional signals. For this, we have applied a Recurrent Neural Network (RNN) with self-attention approach, and a deep audio-to-target end-to-end approach. In addition to these models, we use Support Vector Machines (SVMs), a multimodal Transformer and a fine-tuned NLP Transformer Albert to predict the classes of \emph{\musetopic}.

\vspace{-0.2cm}
\subsection{Early Fusion LSTM-RNN with Self-Attention}
In order to address the sequential nature of the input features, we utilise a Long Short-Term Memory (LSTM)-RNN based architecture. The input feature sequences are input into two parallel LSTM-RNNs with hidden state dimensionality equal to 40, to encode the two corresponding query and value vector sequences. A self-attention sequence is calculated by means of a query and key dot product using a sequence-wide attention window. The attention and query sequences are then concatenated. For the continuous-time tasks \textit{\musewild} and \textit{\musetrust}, the resulting hidden vector for each time step is further encoded by a feed-forward layer that outputs a one-dimensional prediction sequence per prediction target. For the \textit{\musetopic} task, we instead apply global max-pooling, to integrate the sequential information into one hidden state vector, which is then input into a feed-forward layer to provide the logits. In the former case, all the input samples are further segmented into 50 time-step sub-segments which are all used for training, whereas in the latter we pad/crop all sequences to 500 steps.

\vspace{-0.1cm}
\subsection{End-to-End Learning}
As our end-to-end baseline we use End2You~\cite{tzirakis2018end2you}; an open-source toolkit for multimodal profiling by end-to-end deep learning~\cite{tzirakis2017end, tzirakis2019real}. For our purposes, we utilise three modalities, namely, audio, visual, and textual. Our audio model is inspired by a recently proposed emotion recognition model~\cite{tzirakis2018end}, and is comprised of a convolution recurrent neural network (CRNN). In particular, we use 3 convolution layers to extract spatial features from the raw segments. Our visual information is comprised of the \vgg features, where we use zero vectors when the face is not detected in a frame. Finally, as text features we use \ft, where we replicate the text features that span across several segments. We concatenate all uni-modal features and feed them to a one layer LSTM to capture the temporal dynamics in the data before the final prediction. 

\vspace{-0.1cm}
\subsection{Multimodal Transformer}
As baseline for the non-sequential predictions of \musetopic, we choose the Multimodal Transformer (MMT) \cite{yao2019MultT}. By using aligned and unaligned vision, language, and audio features for single label prediction, it outperformed state-of-the-art methods in a more text-focused Multimodal Sentiment Analysis setting. MMT merges multimodal timeseries using a feed-forward fusion process consisting of multiple crossmodal Transformer units. At the core of this network architecture are crossmodal attention modules which fuse multimodal features by directly attending to low-level features across all modalities. To predict topics, valence, and arousal we always utilise 3 feature sets, either of our three (tri), or of only two (bi) different modalities fed into the network. We noticed that after approximately 20 epochs the network converged. The model uses 5 crossmodal attention heads and an initial learning rate of $10^{-3}$.
\vspace{-0.1cm}
\subsection{Albert}
To reflect the current trend towards Transformer language models, such as Bidirectional Encoder Representations from Transformers (BERT) \cite{devlin-etal-2019-bert}, we include one of the latest versions, Albert \cite{Lan2020ALBERT:}, as a purely text-based baseline model. The authors of Albert proposed parameter reduction techniques, so that the total memory consumption is lower while increasing the training speed. These models supposedly scale better than the original BERT. The architecture is able to achieve state-of-the-art results on several benchmarks, despite having a relatively smaller number of parameters. For our purposes, we found a supervised tuning on the train partition for 3 epochs and balanced class weights to have the best effect. We applied a learning rate of $10^{-5}$ for the adjusted Adam Optimiser and set $\epsilon$ to $10^{-8}$. With a sequence length of $300$, the batch size has to be limited to $12$ samples to be trained with 32GB GPU memory. 

\vspace{-0.5em}
\subsection{Support Vector Machines}
For the task of emotion prediction in the Sub-Challenge \musetopic only, we choose also to include results obtained through the use of conventional and easily reproducible Support Vector Machines (SVMs). These experiments employ the \textsc{Scikit-learn} toolkit, with a \textsc{LINEARSVR} classifier. No standardisation or normalisation was applied to any of the reported feature sets. The complexity parameter C was always optimised from $10^{-5}$ to $1$ during the development phase, and the best value for C is reported. In contrast to our other approaches, we retrain the model on a concatenation of the train and development sets to predict the final test set result. 

\vspace{-0.5em}
\section{Baseline Results}

\subsection{\musewild}
We evaluated several feature sets and combinations for the prediction of the continuous arousal and valence (see \Cref{tab:base13} for detailed results). For the prediction of arousal, the LSTM-RNN with self-attention using LLDs as input features, achieved the best result of all applied systems with a CCC of $.3088$ on the devel set, and $.2884$ on the test set. However, the combined metric (mean of valence and arousal) is considerably lower (CCC: $.1931$ on devel) due to the poor efficiency on the prediction of valence. 
Therefore, we define the \textbf{end-to-end framework utilising \ft, \vgg, and audio representations learnt from the raw audio signal as our baseline}, achieving a CCC of $.2431$ (on test) for the prediction of valence and $.2706$ (on test) for the prediction of arousal. We report a combined score of \textbf{$.2568$} (on test) for this system.

\begin{table*}[t!]
\caption{Reporting Arousal, Valence and Combined ($0.5 \cdot Arousal + 0.5 \cdot Valence$) for \textbf{\musewild} and Trustworthiness for \textbf{\musetrust}, both using concordance correlation coefficient (CCC). As feature sets \ft (FT), \egm (Ge), \ds (DS), \go (Go), \vgg (VG), and \xce (X) and all visual (aV) are fed into the models. Furthermore, the raw audio signal (RA) is used in End2You, and \lld (LDD) are utilised for \musetrust in order to predict trustworthiness. All utilised features of \musewild and \musetrust are aligned to the label timestamps by imputing missing values or repeating the word embeddings for \ft. }
% the arousal (Ap) and valence (Vp) predicted signals 

\resizebox{0.6\linewidth}{!}{%linewidth
    \begin{tabular}{ll|ccc|c}
    \toprule
    System      & Features           & \textbf{Valence}      & \textbf{Arousal}      & \textbf{Combined}   & \textbf{Trustworthiness} \\ 
                &                & devel / test          & devel / test          & devel / test          & devel / test\\ \hline
    LSTM + Self-ATT     & LLD                 & .0711 / .0349         & \textbf{.3078 / .2834} & .1894 / .1592 & .2560 / .1343 \\
    LSTM + Self-ATT     & DS                    & .0165 / .0024                   & .1585 / .1723                    &  .0875 / .0874 & .2019 / .1701 \\
    LSTM + Self-ATT     & Ge  & .0435 / -.0097                     & .1090 / .0827                   & .0762 / .0365                    &  .1576 / .1385  \\
    LSTM + Self-ATT     & FT  & .1273 / .1816                     & .0959 / .1074                   & .1116 / .1445                    &  .2278 / .2549  \\
    LSTM + Self-ATT     & Ge + FT  & .0520 / .0361                     & .1375 / .1018                   & .0947 / .0690                    &  .2296 / .2054  \\
    LSTM + Self-ATT     & X  & .0499 / .0426                     & .0776 / .0683                   & .0638 / .0555                    &  .1178 / .1664  \\
    LSTM + Self-ATT     & aV  & .0098 / .0272                     & .1598 / .1227                   & .0848 / .0749                    &  .1167 / .1378  \\
    LSTM + Self-ATT     & Ge + FT + V  & .0393 / .0654                     & .1809 / .0865                   & .1101 / .0760                    &  .1245 / .1695  \\
    End2You & FT + VG + RA                    & \textbf{.1506 / .2431} & .2587 / .2706 &  \textbf{.2047 / .2568}  & \textbf{.3198 / .4128}  \\   
    End2You-Multitask & FT + VG + RA                 & -- & -- &  --  & .3264 / .4119  \\ \bottomrule  
    % End2You & FT + VG + RA + Ap                & -- & -- &  --  & .3318/ .4345  \\   
    % End2You & FT + VG + RA + Vp                & -- & -- &  --  & \textbf{.3450/ .4359}  \\ \bottomrule  
    \end{tabular}
}
\label{tab:base13}
\vspace{-0.2cm}
\end{table*}

\vspace{-0.2cm}
\subsection{\musetopic}
\Cref{tab:base2a} shows the results of the baseline systems on the language-centric task of topic prediction. In line with recent research, the state-of-the-art NLP Transformer \textbf{Albert}, fine-tuned on the training set, achieved with \textbf{$76.79$\,\% (combined, on test) the best baseline result} leaving the second best system, the Multimodal Transformer utilising \ft, \egm, and \fau features ($52.98$\,\% on test), far behind. The most successful configuration of the LSTM + Self-attention, the only not Transformer-based  architecture, has another nearly $15$\,\% performance gap (on test $37.37$\,\%) to the MMT demonstrating the competitiveness of our baseline and the suitability of Transformers for this task.

\begin{table*}[t!]
\caption{\musetopic: Reporting Unweighted Average Recall (UAR), F1, and Combined ($0.66 \cdot F1 + 0.34 \cdot UAR$) for the topic predictions. As feature sets \ft (FT), Raw Text (RT), \egm (eG), \ds (DS), \vgg (VG), \xce (X), \op (OP), \go (Go), \fau (AU) and all visual features (aV) are used. Two types of alignment are used to a) align to \egm (GA), and b) aggregate on \ft word features (FA).}
\resizebox{0.6\linewidth}{!}{ % \columnwidth
\begin{tabular}{lll|ccc}
\toprule
System      & Features          & Alig. & \textbf{F1}       & \textbf{UAR}      & \textbf{Combined} \\ 
            &                   &           & devel / test      & devel / test      & devel / test          \\ \hline   
LSTM + Self-ATT  & DS                & GA          & 19.85 / 34.60 & 12.95 / 35.00  & 17.50 / 34.74       \\
LSTM + Self-ATT  & eG                & GA          & 19.02 / 34.44 & 12.34 / 33.94  & 16.75 / 34.27       \\
LSTM + Self-ATT  & FT                & GA          & 24.62 / 36.19 & 15.25 / 36.22  & 21.44 / 36.20       \\
LSTM + Self-ATT  & eG + FT                & GA  & 20.38 / 35.32 & 13.13 / 34.87  & 17.92 / 35.16       \\
LSTM + Self-ATT  & X               & GA          & 26.06 / 36.83 & 20.42 / 36.61  & 24.14 / 36.75       \\
LSTM + Self-ATT  & aV              & GA          & 27.42 / 34.92 & 21.57 / 34.41  & 25.43 / 34.75       \\
LSTM + Self-ATT  & eG + FT + V   & GA          & 27.58 / 37.14 & 20.08 / 37.14  & 25.03 / 37.14   \\
Fine-tuned Albert  & RT                &  --         & \textbf{71.69} / \textbf{76.59} & \textbf{69.56} / \textbf{77.18}  & \textbf{70.96} / \textbf{76.79}       \\  
MMT         & FT + eG + X       &  --         & 48.24 / 53.18 & 41.49 / 50.44  & 44.86 / 51.81       \\   
MMT         & FT + eG + X       &  FA       & 49.21 / 52.06 & 40.78 / 49.68  & 46.35 / 51.25       \\  
MMT         & FT + eG + VG      &  --         & 44.72 / 50.71 & 35.60 / 45.19  & 41.62 / 48.84       \\    
MMT         & FT + eG + Go      &  --         & 44.72 / 53.73 & 38.14 / 49.85  & 42.48 / 52.41                 \\
MMT         & FT + eG + AU      &  --         & 46.22 / 54.52 & 40.66 / 49.99  & 44.33 / 52.98                 \\  
MMT         & FT + Go + OP      &  --       & 45.17 / 52.30  & 37.82 / 48.61   & 42.67 / 51.05                   \\  
MMT         & FT + DS + Go      &  --         & 44.42 / 52.06 & 36.77 / 49.59  & 41.82 / 51.22                  \\      
\bottomrule   
\end{tabular}
}
\label{tab:base2a}
\end{table*}

\begin{figure}[t!]
    \centering % to be replaced
    \includegraphics[ width=0.98\columnwidth]{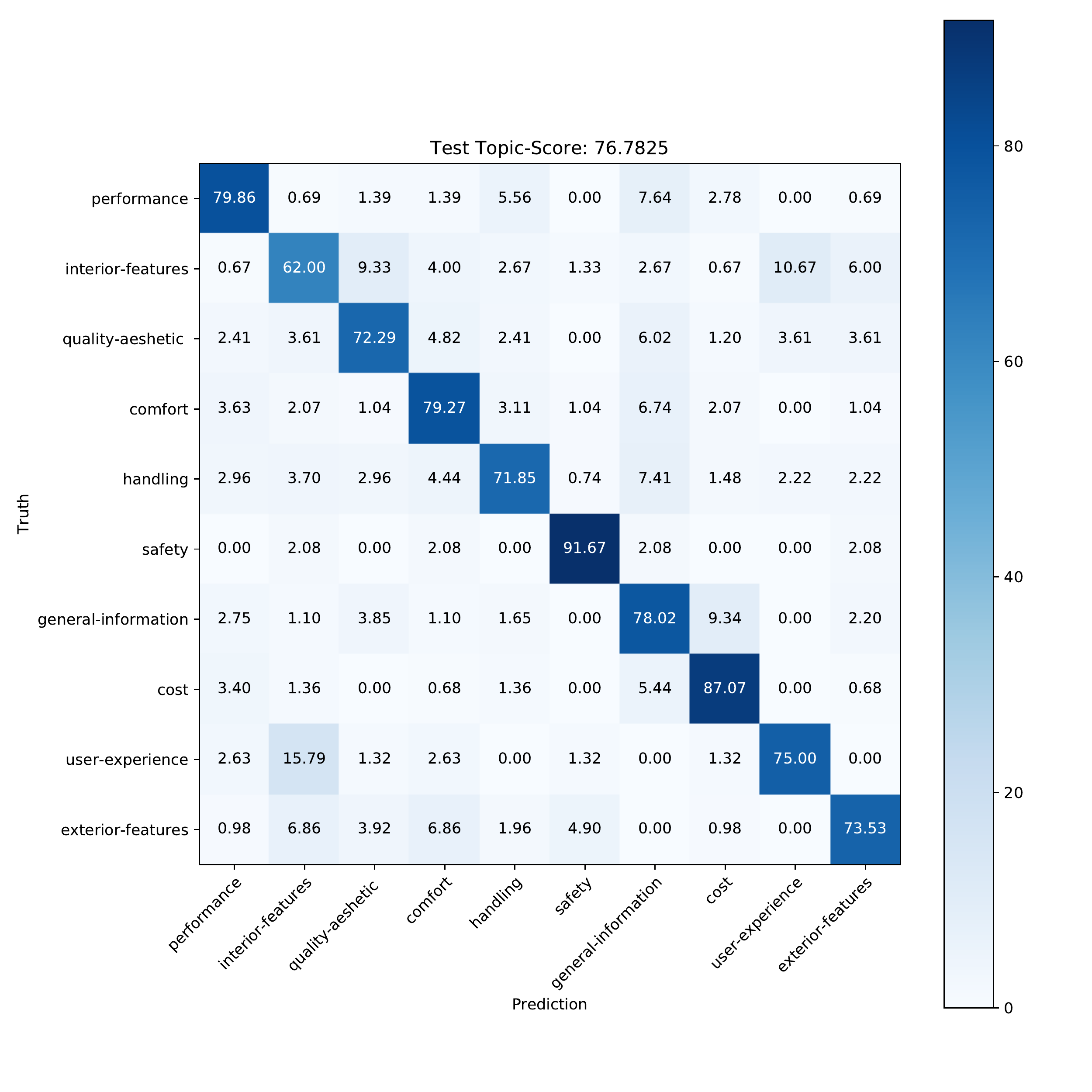}
    % to be replaced by pdf version
    %\includegraphics[width=0.3\linewidth]{figures/arousal_mmt_dsgevg.png} 
    %\includegraphics[width=0.3\linewidth]{figures/XEF_test.png} \\
    \vspace{-1em}
    \caption{Relative confusion matrix over all 10 topics of fine-tuned Albert (left) as well as the MMT (\ft, \egm and \xce) for the prediction of valence (middle) and MMT (\ft, \egm and \xce) for the prediction of arousal (right) classes on the test partition for sub-challenges \musetopic.}
    \vspace{-1em}
    \label{fig:cf}
\end{figure}

For the task of emotion (valence and arousal) prediction in the \musetopic sub-challenge, we also report baseline results in \Cref{tab:base2b}. Here, the picture is more balanced with some system failing to achieve results higher than chance level (33\,\%) on test \eg the fine-tuned Albert. Overall, the \textbf{Multimodal Transformer achieved with $38.81$\% (combined valence and arousal) the best results utilising \ft, \egm, and \xce}. The same configuration is also \textbf{most successful in predicting valence ($40.12$\%) on test}. The utilised SVMs, chosen due to their scalability on high dimensional data, showed results comparable to most state-of-the-art approaches. In particular, for the prediction of arousal, the \textbf{\vgg features result is the best Combined F1 and UAR of $42.67$ on test}. These SVM results lead us to assume that this task may benefit from a more traditional feature-level analysis. The confusion matrix for all tasks are depicted in \Cref{fig:cf}.

\begin{table*}[t!]
\caption{\musetopic: Reporting Unweighted Average Recall (UAR), F1, and Combined ($0.66 \cdot F1 + 0.34 \cdot UAR$) for the 3-class valence and arousal predictions and the combined (mean) of valence and arousal. As feature sets \ft (FT), Raw Text (RT), \egm (eG), \ds (DS), \vgg (VG), \xce (X), \go (Go), \op (OP), \fau (AU), and all visual feature set (aV) are used. Two types of alignment are used to a) align to \egm (GA) or b) aggregate on \ft word features (FA).}
\resizebox{0.9\linewidth}{!}{%
\begin{tabular}{lll|ccc|ccc|c}
\toprule
System      & Features          & Alig. & \multicolumn{3}{c|}{\textbf{c-Valence}}      & \multicolumn{3}{c|}{\textbf{c-Arousal}}   & \textbf{Combined}    \\ 
\multicolumn{3}{l|}{}                        & \textbf{F1} & \textbf{UAR} & \textbf{Combined} & \textbf{F1} & \textbf{UAR} & \textbf{Combined} & \\
            &                   &           & devel / test          & devel / test  & devel / test & devel / test          & devel / test  & devel / test  \\   \hline
Fine-tuned Albert  & RT               &           & 36.18 / 34.21 & 33.17 / 33.05 & 35.16 / 33.81 & 
33.33 / 37.14 & 33.69 / 34.30 & 33.45 / 36.18  
& 34.30 / 35.00 \\
%LSTM + ATT  & FT                &           & & &               \\
LSTM + Self-ATT  & DS                &      GA     & 34.17 / 34.60 & 34.07 / 35.00 & 34.13 / 34.74 & 38.03 / 37.54 & 38.43 / 36.78 & 38.17 / 37.28 & 36.15 / 36.01         \\
LSTM + Self-ATT  & eG                &      GA     & 33.26 / 34.44 & 32.16 / 33.94 & 32.89 / 34.27 & 34.39 / 33.33 & 34.44 / 32.87 & 34.41 / 33.18 & 33.65 / 33.73        \\
LSTM + Self-ATT  & FT                &      GA     & 38.41 / 36.19 & 37.75 / 36.22 & 38.18 / 36.20 & 35.15 / 34.92 & 35.78 / 37.10 & 35.37 / 35.66 &36.78 / 35.93         \\
LSTM + Self-ATT  & eG + FT           &      GA     & 34.92 / 35.32 & 34.05 / 34.87 & 34.63 / 35.16 & 34.39 / 35.48 & 34.48 / 35.42 & 34.42 / 35.46 &34.53 / 35.31             \\
LSTM + Self-ATT  & X                 &      GA     & 36.21 / 36.83 & 35.75 / 36.61 & 36.06 / 36.75 & 40.38 / 35.16 & 40.51 / 34.87 & 40.43 / 35.06   & 38.24 / 35.91         \\
LSTM + Self-ATT  & aV               &      GA     & 35.61 / 34.92 & 35.10 / 34.41 & 35.44 / 34.75 & 38.11 / 34.21 & 38.26 / 35.39 & 38.16 / 34.61  &36.80 / 34.68            \\
LSTM + Self-ATT  & eG + FT + aV &   GA     & 36.06 / 37.14 & 35.20 / 37.14 & 35.77 / 37.14 & 39.92 / 35.16 & 40.44 / 34.76 & 40.10 / 35.02  &37.93 / 36.08            \\
MMT              & FT + eG + X           &           & 38.28 / \textbf{39.92} & 37.62 / \textbf{40.52} & 38.06 / \textbf{40.12} & 41.87 / 37.30 & 40.83 / 37.87 & 41.52 / 37.50  &39.79 / \textbf{38.81}\\
MMT         & FT + eG + VG          &           & 37.38 / 32.78 & 38.19 / 32.53 & 37.65 / 32.69 & \textbf{47.12} / 41.19 & \textbf{45.55} / 39.01 & \textbf{46.58} / 40.45 &42.12 / 36.57\\
MMT         & DS + eG + VG          &           & 39.40 / 32.54 & 38.08 / 32.40 & 38.95 / 32.49 & 45.77 / 41.03 & 44.66 / \textbf{40.63} & 45.39 / 40.89  &\textbf{42.17} / 36.69\\
MMT         & X  + eG + VG          &           & 38.28 / 36.43 & 37.76 / 37.39 & 38.10 / 36.76 & 45.24 / 40.95 & 43.81 / 38.66 & 44.76 / 40.17&41.43 / 38.46\\
MMT         & FT + eG + AU          &           & 36.93 / 39.92 & 37.35 / 39.57 & 37.07 / 39.80 & 43.15 / 34.76 & 41.88 / 34.87 & 42.72 / 34.80  &39.89 / 37.30\\   
MMT         & FT + eG + OP          &           & \textbf{39.48} / 38.81 & \textbf{39.17} / 38.64 & \textbf{39.37} / 38.75 & 38.88 / 37.70 & 38.95 / 38.10 & 38.90 / 37.83  &39.14 / 38.29\\ 
MMT         & OP + eG + AU          &           & 37.30 / 36.67 & 36.34 / 37.45 & 36.97 / 36.93 & 43.15 / 34.68    & 42.01 / 35.69 & 42.76 / 35.03 &39.87 / 35.98 \\ 
End2You     & FT + eG + X               &           & 37.19 / 33.54 & 35.70 / 33.18 &  36.68 / 33.42  &42.76 / 32.45      & 42.67 / 33.34 & 42.73 / 32.75 & 39.70 / 33.08\\ 
SVM         & eG                    &           & 36.33 / 33.10	& 34.79 / 34.13	&35.81 /	33.45	&43.52 / 34.37	& 42.27 / 33.43	&43.10 /	34.05  &39.45 / 33.75\\ 
SVM         & DS                    &           & 34.08 / 34.29	& 33.21 / 34.07	&33.79 /	34.21	&41.35 / 42.30	&40.18 / 40.18	&40.18 /	41.83 &36.98 / 38.02\\ 
SVM         & X                     &           & 38.28 / 37.94	& 37.09 / 37.94	&37.87 /	37.94	&46.22 / 41.35	&45.25 / 40.52	&45.89 /	41.07 & 41.88 / 39.50\\
SVM         & VG                    &           & 37.08 / 32.94	& 37.01 / 32.63	&37.06 /	32.83	&46.44 / \textbf{42.46}	& 45.21 / \textbf{43.07} & 46.02 /	\textbf{42.67} & 41.54 / 37.75 \\
SVM         & FT                    &           & 37.90 / 36.43	& 36.00 / 35.37	&37.26 /	36.07	&45.17 / 38.25	& 44.53 /	39.67	& 44.95 / 38.74 & 41.10 / 37.40 \\
\bottomrule   

\end{tabular}
}
\label{tab:base2b}
\end{table*}

\vspace{-0.5em}

\subsection{\musetrust}
The results for the prediction of trustworthiness are depicted in \Cref{tab:base13}. Similar to \musewild, the end-to-end baseline system using \ft, \vgg, and raw audio signals gave the best results with $.4128$ CCC on test. The results may improve if the valence and arousal predicted signals are incorporated during training. This can be accomplished in three ways: i) the model from \musewild is utilised to predict arousal and valence on \musetrust; ii) the arousal and valence models can be retrained on \musetrust (we provide train and devel labels); or iii) all three are predicted in a multitask-fashion (one model, 3 outputs) on train and devel, and only trustworthiness is predicted on test. We decided for option (iii). Adding these signals to the \textbf{end-to-end baseline system}, the predictive power of the model is similar to the previous one with CCC $.3264$ on the development set and $.4119$ on the test set. 

\section{Conclusions}
In this paper, we introduced MuSe 2020 -- the first Multimodal Sentiment Analysis in real media assessment challenge. MuSe 2020 utilises the \musec multimodal corpus of emotional car reviews and comprises three Sub-challenges: i) \musewild, where the level of the affective dimensions of valence (corresponding to sentiment) and arousal has to be predicted from a ca.\ 35 hour data subset; ii) \musetopic, where the domain-related conversational topic (10 classes) as well as three classes (low, medium and high) of valence and arousal have to be predicted from video parts containing the discussed topic; and, iii) \musetrust, where the level of continuous trustworthiness has to be predicted from features and/or affective annotations. By intention, we decided to use open-source software to extract a wide range of feature sets to deliver the highest possible transparency and realism for the baselines. Besides the features, we also share the raw data and the developed code for our baselines on a public platform. Results indicate that: i) the level of affection in-the-wild is best predicted when the system is trained on the the raw audio features; ii) for \musetopic, (NLP-specific) Transformers are clearly superior when it comes to the prediction of topics, and no system is clearly outperforming on the three class valence and arousal prediction; and iii), in \musetrust, adding valence and arousal contours as `signals' in addition to other features is beneficial for the prediction of trustworthiness. The baselines also show the challenge ahead in mastering multimodal sentiment analysis, in particular when data are collected in user-generated, noisy environments. In the participants' and future efforts, we expect novel exciting combinations of the modalities -- potentially also such as linking modalities on earlier stages or more closely.
            
\vspace{.2cm}
\section{Acknowledgments}
This project has received funding from the European Union's Horizon 2020 research and innovation programme under grant agreement No.\,115902 (RADAR CNS) and No.\,826506 (sustAGE), 
the EPSRC Grant No.\,2021037, and the Bavarian State Ministry of Education, Science and the Arts in the framework of the Centre Digitisation.Bavaria (ZD.B).
We thank the sponsors of the Challenge BMW Group and audEERING.


\footnotesize
\bibliographystyle{ACM-Reference-Format}
\bibliography{sample-base}


\end{document}